\documentclass[sn-nature,hidelinks]{sn-jnl}

\newcommand{\gtwo}{$g^{(2)}(\tau) $}

\usepackage{graphicx}%
\usepackage{multirow}%
\usepackage{amsmath,amssymb,amsfonts}%
\usepackage{amsthm}%
\usepackage{mathrsfs}%
\usepackage[title]{appendix}%
\usepackage{xcolor}%
\usepackage{textcomp}%
\usepackage{manyfoot}%
\usepackage{booktabs}%
\usepackage{algorithm}%
\usepackage{algorithmicx}%
\usepackage{algpseudocode}%
\usepackage{listings}%
\usepackage{listings}%
\usepackage{newfloat}%
\usepackage{textgreek}%
\DeclareFloatingEnvironment[name={Fig. S}]{suppfigure}%

\hypersetup{
    colorlinks,
    linkcolor = {red!50!black},
    citecolor = {blue!50!black},
    urlcolor = {blue!80!black}
}

\begin{document}

\title[Version 0.1]{Advanced analysis of single-molecule spectroscopic data}

\author[1,4]{Joshua L. Botha}

\author[1,2,3,5]{Bertus van Heerden}

\author*[1,2,3,6]{Tjaart P.J. Kr\"uger}\email{tjaart.kruger@up.ac.za}

\affil[1]{\orgdiv{Department of Physics}, \orgname{University of Pretoria}, 
\orgaddress{\street{Lynnwood Road}, \city{Pretoria}, \postcode{0002},
\state{Gauteng}, \country{South Africa}}}

\affil[2]{\orgdiv{Forestry and Agricultural Biotechnology Institute (FABI)}, 
\orgname{University of Pretoria}, \orgaddress{\street{Lynnwood Road}, 
\city{Pretoria}, \postcode{0002},
\state{Gauteng}, \country{South Africa}}}

\affil[3]{\orgname{National Institute of Theoretical and Computational Sciences (NITheCS)}, \country{South Africa}}

\affil[4]{ORCID: 0000-0002-4268-5517}
\affil[5]{ORCID: 0000-0002-6769-0697}
\affil[6]{ORCID: 0000-0002-0801-6512}
\date{\today}

\abstract{We present \textit{Full SMS}, a multipurpose graphical user interface
(GUI)-based software package for analysing single-molecule spectroscopy (SMS)
data. SMS typically delivers multiparameter data --- such
as fluorescence brightness, lifetime, and spectra --- of molecular- or nanometre-scale particles such as single dye molecules, quantum dots, or fluorescently labelled biological macromolecules. \textit{Full SMS} allows an unbiased statistical analysis of fluorescence brightness through level resolution and clustering, analysis of fluorescence lifetimes through decay fitting, as well as the calculation of second-order correlation functions and the display of fluorescence spectra and raster-scan images. Additional features include extensive data filtering options, a custom HDF5-based file format, and flexible data export options. The software is open source and written in Python but GUI-based so it may be used without any programming knowledge. A multi-process architecture was employed for computational efficiency. The software is also designed to be easily extendable to include additional import data types and analysis capabilities.}


\maketitle

\section{\label{sec:intro} Introduction}

Since the first single-molecule spectroscopy (SMS) measurement in
1989~\cite{Moerner1989b}, the technique has developed rapidly. Also called
``single-particle spectroscopy" when not applied to strictly one molecule, SMS
is applicable to nanoscale systems containing one or a small number of
quantum emitters. The exquisite level of selectivity and sensitivity
of SMS techniques gives access to numerous properties of nanoscale emitters
that are usually masked in conventional ensemble-averaging techniques, thereby
offering a broad scope of applications in physics, chemistry, and
biology~\cite{Moerner2002}. For example, time-dependent SMS measurements
provide information about the dynamics and kinetics of nanoscale systems and can
resolve time-dependent processes without the need to synchronise these processes
in a large collection of identical molecules. The biological applications of SMS
are particularly vast and include observations of protein conformational
dynamics~\cite{Mazal2019}, enzyme reactions~\cite{Ha1999,Jiang2011},
transcription in single DNA molecules~\cite{Kapanidis2006,Hou2020},
light-harvesting complexes switching between different functional
states~\cite{Kruger2017,Kondo2017,Gwizdala2018,Gruber2018}, changes in the
oligomeric states of macromolecules~\cite{Tuson2015}, and changes in the
diffusive states of nanoparticles~\cite{Welsher2014}. Real-time 3D tracking of
biological (macro)molecules allows SMS to be performed with high spatiotemporal
resolution in live cells~\cite{vanHeerden2022}. Applications in non-biological
settings include the use of quantum dots~\cite{Garcia2021}, single organic
molecules~\cite{Toninelli2021}, and nitrogen-vacancy centres in
diamond~\cite{Zhang2021} in photonic quantum technologies and quantum sensing.

Two typical features exhibited in the photon emission from individual nanoscale
emitters are photon antibunching and fluorescence intermittency. These phenomena
are used to verify the presence of a single quantum emitter but have also found
interesting applications. Photon antibunching can be used to count the number of
fluorescent labels in confocal microscopy, investigate multiple coupled emitters
(by analysing higher-order photon correlations), and quantify photophysical
processes through the width of the antibunching dip~\cite{Lubin2022}. While
fluorescence intermittency (often called `blinking') tends to be a hindrance in
most applications, it provides a source of information to improve the spatial
resolution beyond the diffraction limit~\cite{Heilemann2008, Linde2011} and, for
light-harvesting complexes, is a useful means to study
photoprotection~\cite{Kruger2012,Gwizdala2016}.

SMS is a powerful investigation technique, as the measured data contains a rich
amount of information. Numerous approaches have been developed during the past
three decades to give access to diverse types of dynamic information about
various kinds of nanoscaled systems. However, since the developments have been
done across different research groups, each measurement type often requires a
different data analysis program, which necessitates the analysis of each
measurement type to take place typically in isolation from the rest. In
addition, different software packages usually require different file formats and
often lack versatility, having been designed for a particular application. As a
result, it usually takes a great amount of effort to obtain a multiparameter SMS
data set through the combination of different existing software packages.

Although numerous open-source programs give access to their scripts, it is
cumbersome to adapt those scripts to enable other applications and is only a
viable solution for experienced programmers. Commercial software is also limited
in its application, incompatible with other hardware types, and expensive. 

There is, therefore, a need for generalised, open-access analysis suites to
facilitate and standardise the analysis of multiparameter SMS data. Some work in
developing such software has been done, and examples include the Multiparameter
Fluorescence Detection (MFD) software from the group of
Seidel~\cite{Seidel2024}, \textit{PAM} from the group of
Lamb~\cite{Schrimpf2018}, and a change-point analysis toolbox from the group of
Koenderink~\cite{Palstra2021}. A general-purpose file format for SMS and other
photon-by-photon data has also been developed, namely
Photon-HDF5~\cite{Ingargiola2016}. 

Here, we present \textit{Full SMS}, a single workspace where data from multiple concurrent
measurements of the same subject can be viewed and analysed. Unlike the MFD and \textit{PAM}
software mentioned above, our software is largely based on the
analysis of intensity time traces using a statistically robust change-point analysis (CPA). 
CPA, along with subsequent clustering of the intensity levels and
fluorescence lifetime fitting, is a powerful approach to analysing fluorescence
intermittency and identifying the different emissive states of nanoscale
emitters. \textit{Full SMS} is also entirely written in Python and thus is free
to use and extend, setting it apart from MFD, which is closed source, while
\textit{PAM} is open source but written in MATLAB, which is proprietary
software. Unlike the toolbox developed by Palstra and Koenderink, our software
is based on a graphical user interface (GUI). This has several advantages over
simple scripts or command-line programs, the main ones being that easy visual
exploration of the data is enabled and that users need very little technical
knowledge to get started using the software. In this regard, our software is
similar to \textit{Glotaran}~\cite{Snellenburg2012, vanStokkum2023}, a widely
used tool for analysing time-resolved spectra. Additional advantages of
\textit{Full SMS} over the toolbox of Palstra and Koenderink are the display of
raster-scan images and spectral time traces, and automated trimming of traces as
well as built-in data filtering based on intensity or lifetime distributions. We
also developed a custom file format based on the HDF5 format and similar to
Photon-HDF5, but which includes specific additional measurement capabilities
suitable to measurements of single, immobilised particles.

\section{\label{sec:application} Application Features}

\textit{Full SMS} is a GUI-based application that allows users to open, view,
and perform analyses of multiple types of SMS measurements. Written in Python,
it uses a Python binding of the Qt software suite for the GUI functionality,
which produces a cross-platform application. A multi-process architecture was
used to allow for parallel computation operations, which is notoriously
challenging in Python due to the so-called global interpreter lock, but it is
crucial to allow for the quick analysis of large data sets. The measurements are
stored in the HDF5 format, developed by the HDF Group, which allows for
efficient storage, a hierarchical-relational grouping of measurements, and rich
meta-data capabilities.

\textit{Full SMS} is designed to analyse datasets in a single HDF5 file
containing multiple \textit{groups} of data and metadata, each group
corresponding to a measurement of a single particle. Although the specific
structure of the file is custom to \textit{Full SMS}, the raw data can be read
using any tool designed for HDF5 files. All the analyses described below as well
as subsequent exporting of the results can be performed fully within the GUI,
with no programming knowledge needed from the user.

Fig.~S1 (Supplementary Information) shows a screenshot of the main window of the
GUI. On the left, individual particles can be selected. At the top, tabs can be
selected for each major datatype or analysis. Once a dataset is loaded, the
lower left pane displays a description as entered by the user at the time of the
measurement or added to the HDF5 file.

\subsection{\label{sec:analysis} Analysis}

We first give a summary of the analyses and tools that the software provides and
follow this with a discussion of the main features, using mainly the data from one 
measured particle as a step-wise illustration of the analysis capabilities.

Many kinds of analyses can be performed on SMS data, depending on the type(s) of
measurements made. A predominant SMS measurement modality is fluorescence
intensity (also called `brightness'), either directly by measuring the photon
flux within predetermined time windows or indirectly by measuring photon arrival
times and then calculating the intensity in time bins (see Section
\ref{sec:intensity-level-resolution}). The latter, called time-correlated
single-photon counting (TCSPC), typically provides a statistical distribution of
photon arrival times relative to the moment of excitation of the sample by a
pulsed light source, a method commonly known as time-tagged time-resolved (TTTR)
data collection. Further analysis allows the resolved intensity levels to be
grouped by employing a clustering algorithm, and the statistically most probable
number of states to be determined (see Section
\ref{sec:intensity-level-grouping}). \textit{Full SMS} additionally enables globally grouping over the resolved intensity levels from multiple measurements to resolve underlying states with greater confidence (see Section \ref{sec:intensity-level-grouping}). \textit{Full SMS} can analyse TCSPC data to
extract fluorescence lifetimes for the whole measurement, but also for each
resolved intensity level or intensity group (see Section
\ref{sec:fluorescence-decay-fitting}). It can also calculate the second-order
photon correlation function \gtwo (see Section
\ref{sec:second-order-correlation}).

Additional features, available in separate tabs, are data filtering options (see Section~\ref{sec:data-filtering}), the
display of measured fluorescence spectral sequences and raster-scan images, and
options for exporting the analysed data (see
Sections~\ref{sec:spectra}--\ref{sec:export}). Raster-scan images are useful for
the spatial mapping of the measured particles in concurrent imaging modalities
or simply to locate individual particles before measurement.

The software also provides several other features (see Section \ref{sec:tools}),
such as data format conversion, photon burst detection, and the capability to define regions of interest for each measurement individually, either manually or procedurally, by defining intensity and temporal thresholds.

\subsubsection{\label{sec:intensity-level-resolution} Intensity Level Resolution}

Analysis of an emitter's intensity trace is useful not only to assess its
intrinsic brightness but especially to investigate time-dependent processes
giving rise to fluorescence intensity fluctuations, which may result from static
or dynamic quenching, photobleaching, or other photophysical changes. Amongst
these, fluorescence intermittency, a telltale sign of a single emitter, is evidenced
by abrupt and reversible decreases in the photon emission rate for periods
anywhere between sub-milliseconds to tens of minutes \cite{Basche1998, Frantsuzov2008}. 
To correctly analyse intensity fluctuations, the points at which
the intensity changes occur must be identified. An additional benefit of such a capability
is to filter out unwanted segments of the intensity trace, for example, those corresponding to photobleaching.

\begin{figure}[htp]
    \centering
    \includegraphics[width=1\linewidth]{
    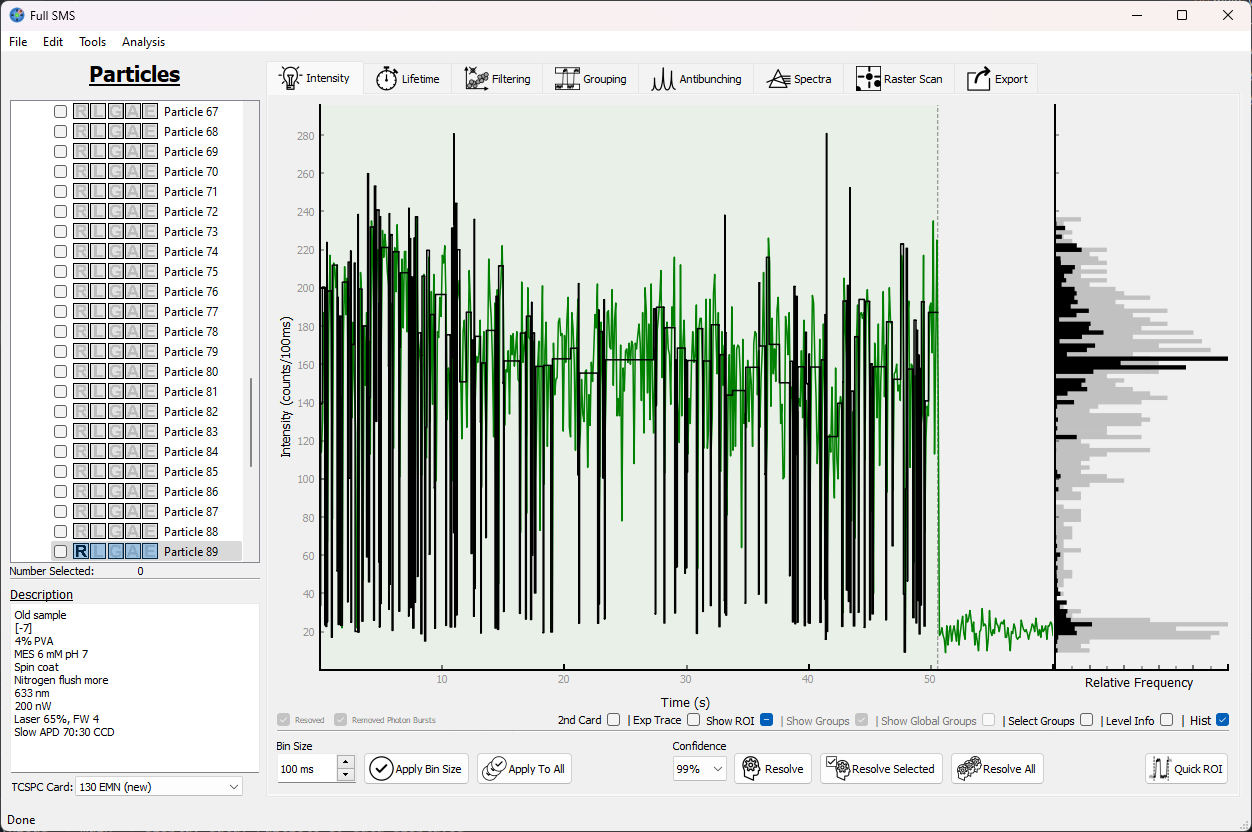}
    \caption{
        Screenshot of a resolved fluorescence brightness trace of a single 
        Alexa Fluor 647 dye molecule. See text for details.
        }
    \label{fig:gui_resolved_trace}
\end{figure}

Identifying the points at which a change in the photon statistics occurs is
typically done using a change-point analysis (CPA)~\cite{Watkins2005,
Ensign2010}. This can be done using time-binned intensity traces or directly
using the photon arrival times. Working in the intensity regime (the first
approach) is often done to simplify the analysis workflow, and suitable CPA
algorithms have been developed~\cite{Kruger2011a, Song2017, Li2019}. The choice
of bin size is crucial in this case as it determines the temporal resolution
with which dynamics can be resolved. However, data binning inevitably introduces
a bias in the analysis~\cite{Watkins2005, Hoogenboom2006, Ensign2010,
Houel2015}. This bias is especially relevant when the dynamics of the intensity
fluctuations occur at a rate comparable to the chosen bin size.

Change-point detection in the temporal regime avoids the bias introduced by data
binning by working with the individual photon arrival times. This enables the
optimal amount of information retrieved from a photon stream, limited only by
the shot noise. We opted for the CPA method introduced in Ref.
\cite{Watkins2005} because of its independence of a physical model underlying
the photon statistics. This method compares the duration between each pair of
consecutive photon detections to the durations between neighbouring pairs of
photon detections and assigns a change point to the instant when a sustained
abrupt change in these durations occurs. A chosen confidence value determines a
statistical criterion for each potential change point that must be met before
acceptance. The higher the confidence value, the more stringent the criteria
that need to be met, and the less sensitive the change-point detection becomes.
In addition, each change point has an associated temporal error region, defined
by the first and last detected photons that pass the statistical test. The
resulting change points separate the measurement into segments.

After loading a measured data set, an intensity trace is displayed in the first
tab of \textit{Full SMS} with a corresponding intensity histogram on the right
(see Fig.~S2). The intensities are binned as
defined by the user's chosen bin size. This binning is only used for the visual 
representation of the intensity and can be changed at any time without impacting 
the rest of the analysis). Once loaded, the intensity levels are resolved 
automatically unless the operation is aborted by the user. The
screenshot in Fig.~\ref{fig:gui_resolved_trace} is an example of a measurement
of single Alexa Fluor 647 (hereafter called Alexa) dye molecules --- a common dye
used for microscopic chromophore labelling --- that was subjected to CPA. The
green trace represents the binned intensities calculated from the measured time
series, whereas the black trace represents the resolved levels. It is worth
noting that the CPA intensity level resolution is independent of the bin size
used for the intensity representation, as the CPA operates on the underlying
time series. The pane on the right shows a histogram of the binned intensities
(grey) and a level-duration weighted distribution of the resolved intensity
levels (black). Below the measurement window, the bin size for display as well
as the confidence level of the CPA analysis can be set, options are given to run
the CPA on the current, selected, or all particles, and the region-of-interest
(ROI) functionality can be accessed (see Section~\ref{sec:trace-roi}). More
information is available about the whole trace
(Fig.~S3), or a selected level
(Fig.~S4), by selecting \textit{Level
Info}. Finally, an additional functionality (activated through \textit{2nd Card}) 
is the display of two-channel data (Fig.~S5). Currently, 
the two channels are analysed separately, except for second-order correlation 
analysis (see Section~\ref{sec:second-order-correlation}).

\subsubsection{\label{sec:fluorescence-decay-fitting} Fluorescence Decay Fitting}

\begin{figure}[htp]
    \centering
    \includegraphics[width=1\linewidth]{
    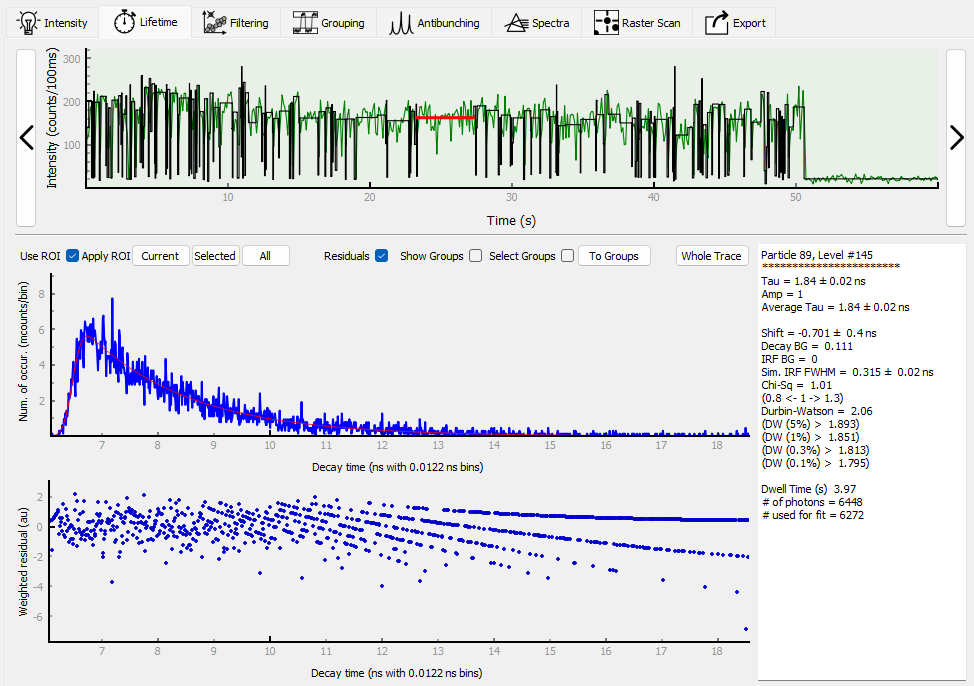}
    \caption{
        Screenshot of an example lifetime fit corresponding to a single resolved 
        brightness level of the Alexa trace shown in 
        Fig.~\ref{fig:gui_resolved_trace}. Fitting residuals are shown in the 
        lowest pane and fitting results are on the right. See text for details.
    }
    \label{fig:gui_lifetime_fit_single_level}
\end{figure}

To extract fluorescence lifetime information, the photon arrival times relative
to the excitation laser pulse are histogrammed and fitted in an iterative
reconvolution minimisation procedure, which is based on either the least-squares
(LS)~\cite{Grinvald1974} or maximum likelihood (ML)~\cite{Bajzer1991}
approaches. The approaches give identical results for high photon counts, and in
such cases LS is preferred since it is more computationally efficient and
robust. For low photon counts, ML should be used since it correctly accounts for
Poisson noise~\cite{Bajzer1991}. Lifetime fitting is performed within the
\textit{Lifetime} tab of the software
(Fig.~\ref{fig:gui_lifetime_fit_single_level}). A fit can be performed on an
entire experimental trace or individual resolved levels using only the photons
from the level. After performing the grouping, a fit can also be done for each
level group, as each intensity state is typically associated with a specific
lifetime. This is especially powerful for dim levels or data with very fast
switching where the individually fitted levels do not contain many photons from
which to extract a lifetime. 

\begin{figure}[htp]
    \centering
    \includegraphics[width=0.8\linewidth]{
    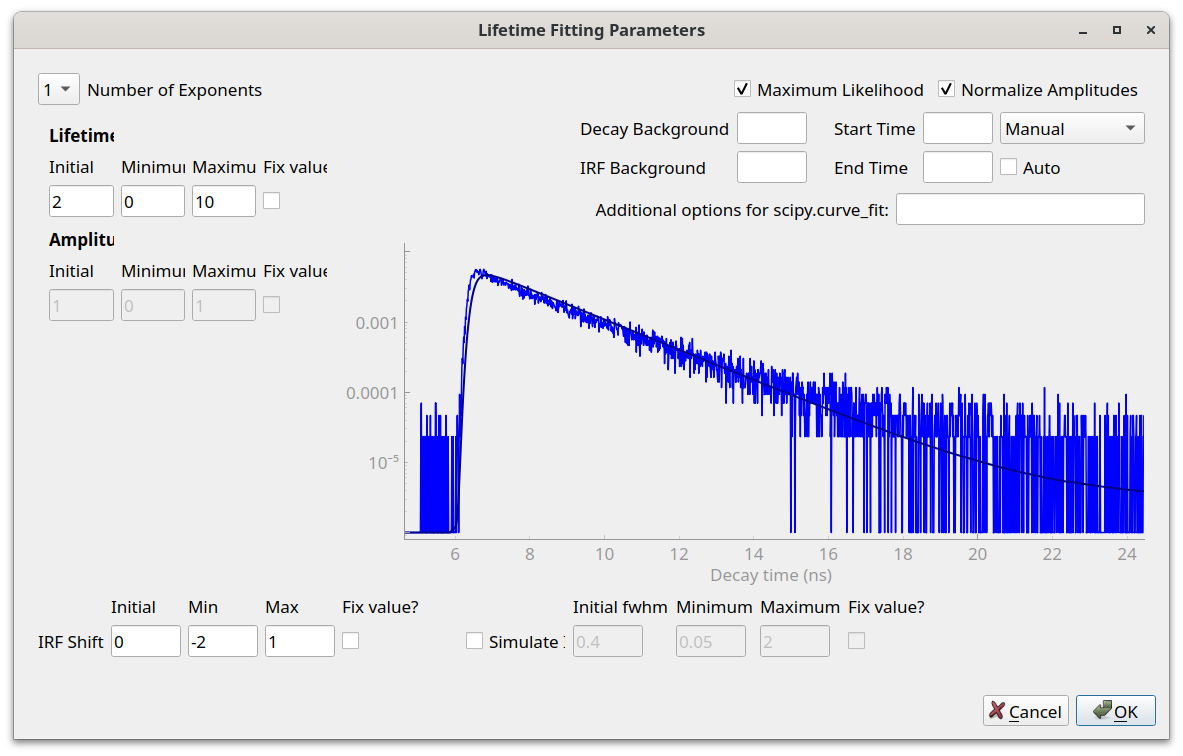}
    \caption{
        Screenshot of the user-definable parameters for the lifetime fitting,
        showing the decay histogram for the Alexa trace in Fig.~\ref{fig:gui_resolved_trace}.
        At the top left is the number of lifetime exponents that should be fit.
        Below that, initial values and boundaries for the lifetimes can be provided.
        On the bottom left, the initial value and boundaries for the IRF shift can be defined.
        The user has the option for an IRF to be simulated, which can often result in better fits.
        Additional parameters can be defined on the top right, namely background values for the 
        decay and IRF and the start and end times of the fit window to be used. These values can 
        also be set automatically by the software. Lastly, at the top right are options to use 
        maximum-likelihood fitting and to normalise the lifetime amplitudes.
        }
    \label{fig:gui_lifetime_fitting_parameters}
\end{figure}

The intensity trace is displayed at the top of the \textit{Lifetime} tab
(Fig.~\ref{fig:gui_lifetime_fit_single_level}), and individual levels or groups
can be selected by clicking on them. Below the displayed intensity trace are
options to apply the trace ROI for the current, selected, or all particles
(i.e., to include only the ROI photons in the lifetime fits), the option to show
the fit residuals plot (which is enabled in
Fig~\ref{fig:gui_lifetime_fit_single_level}), and options to show the groups
instead of the resolved levels in the intensity trace plot, so that they can be
selected and their decay histograms and fitting results viewed. An interactive
dialog is used to choose the parameters for the fitting
(Fig.~\ref{fig:gui_lifetime_fitting_parameters}). After performing the fit(s),
the results are shown in the \textit{Lifetime} tab
(Fig~\ref{fig:gui_lifetime_fit_single_level}). The fits are shown along with the
residuals, if enabled, and the fitting parameters and goodness-of-fit statistics
are shown in the right pane. To evaluate the goodness-of-fit, we use the
reduced $\chi^2$ value (only for LS) and the Durbin--Watson (DW)
parameter~\cite{Durbin1951} (for both LS and ML). The latter measures
autocorrelation in the residuals, which is more sensitive to minor fitting
errors than $\chi^2$. Our code also provides tools to automatically identify
suitable boundaries for curve-fitting, which is important for evaluating the fit
using the DW parameter.

\subsubsection{\label{sec:data-filtering} Data Filtering}

\begin{figure}[htp] 
    \centering 
    \includegraphics[width=1\linewidth]{
        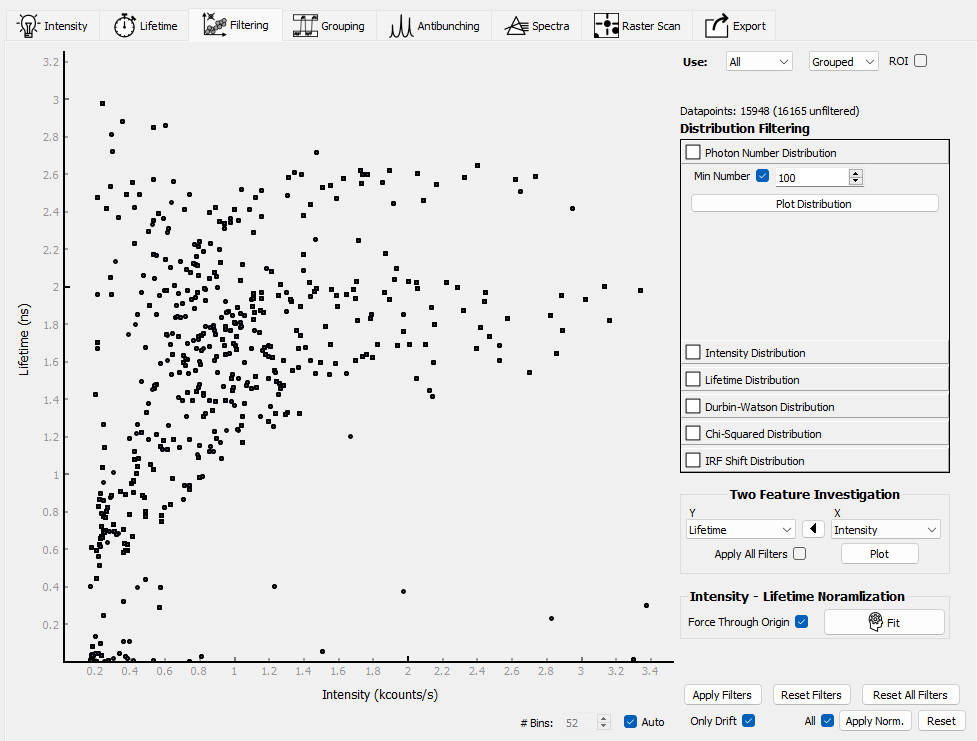}
    \caption{
    Data filtering example for Alexa, showing a scatter plot of intensity and
    lifetime values for grouped levels. The data is filtered to include only
    groups with more than 100 photons. The various filtering criteria are
    mentioned in the main text. The number of histogram bins is set automatically 
    but can also be set manually under \textit{\#Bins} by first unselecting 
    \textit{Auto}, and options for intensity--lifetime normalisation. If
    \textit{Only Drift} is checked, only intensities below the regression line 
    will be normalised, else all intensities are normalised. Selecting 
    \textit{All} applies the normalisation to all levels and groups instead of 
    only the currently displayed data.}
    \label{fig:gui_filtering}
\end{figure}

In the \textit{Filtering} tab of the software (Fig.~\ref{fig:gui_filtering}),
the data can be filtered after level resolution, grouping, and lifetime fitting.
At the top right are options to use all or selected particles or only the
current particle, to use either resolved or grouped levels, and to include only
levels and groups within the ROI of each trace. Below this, the total number of
datapoints after and before filtering is shown. The parameters used for
filtering are the number of photons, intensity, average fitted lifetime, DW
parameter of the fit, $\chi^2$ of the fit, and fitted IRF shift for each level
or group. A histogram of each parameter or a 2D scatter plot of two filter
values can be plotted to aid in determining appropriate filter parameters (as
shown in Fig.~\ref{fig:gui_filtering}). Examples include filtering out levels
with less than a certain number of photons (e.g., 100) or with lifetime or
intensity values that are outliers. To plot a histogram for a specific
parameter, \textit{Plot Distribution} can be selected under the relevant
parameter option under \textit{Distribution Filtering}. A scatter plot can be
viewed by selecting the relevant two parameters under \textit{Two Feature
Investigation} and selecting \textit{Plot}. If an intensity--lifetime scatter
plot is made, a linear regression fit can be performed for the data by clicking
\textit{Fit} under \textit{Intensity - Lifetime Normalisation}, with an option
of forcing the fit through the origin. Once this is done, a normalisation can be
applied by clicking \textit{Apply Norm}. This changes the level intensities to
match their lifetimes based on the regression fit.  

\subsubsection{\label{sec:intensity-level-grouping} Intensity-Level Grouping}

The information gained from CPA can be significantly increased by statistically
associating the resolved intensity levels with one another. A grouping or clustering of this kind has
multiple uses. In the first place, it serves as a tool to resolve underlying
states in the system being studied~\cite{Watkins2005}. An added advantage is
performing fluorescence decay fitting of levels with low numbers of detected
photons. The efficacy of fluorescence decay fitting (discussed in Section
\ref{sec:fluorescence-decay-fitting}) depends, amongst other factors, on the
number of data points used. Investigating states that emit very few photons,
therefore, often proves to be problematic. By grouping several of these dark or
dim states, a successful fit is possible where it would otherwise involve
a large uncertainty.

\begin{figure}[htp]
    \centering
    \includegraphics[width=0.9\linewidth]{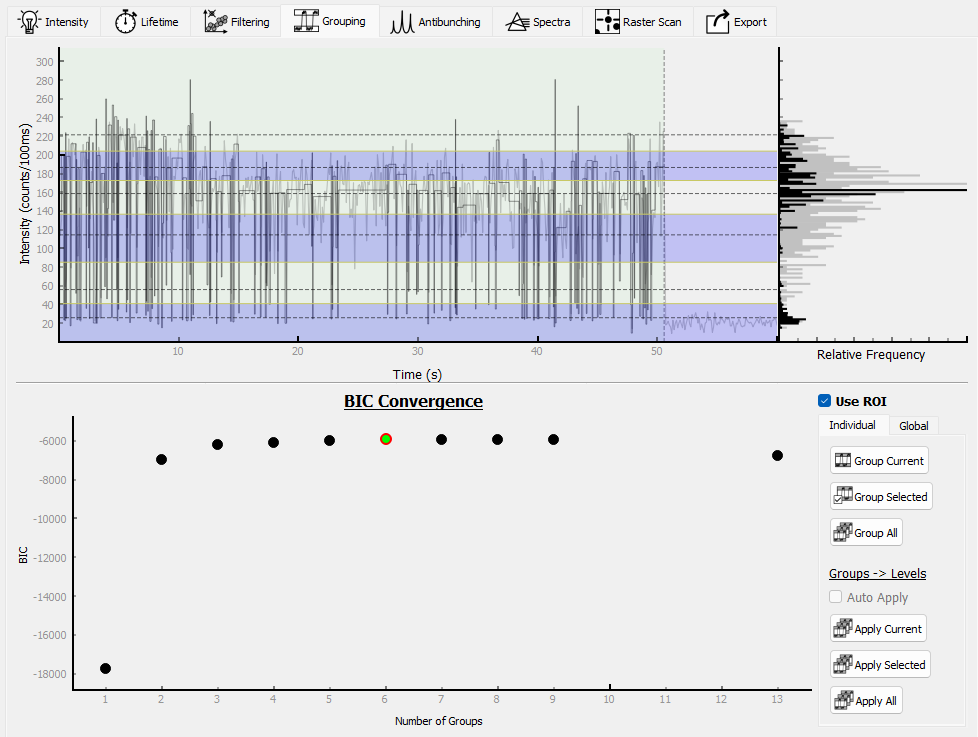}
    \caption{
        Screenshot of an intensity-level grouping of  the Alexa intensity trace 
        in Fig.~\ref{fig:gui_resolved_trace}. See text for details.
    }
    \label{fig:gui_grouping_example}
\end{figure}

The clustering method implemented in \textit{Full SMS} is a mixed model of an
agglomerative hierarchical clustering (AHC) method and expectation maximisation
(EM) clustering method~\cite{Ward1963, Watkins2005, Li2019}. AHC starts by
considering each intensity level as a group of its own. It then merges the pairs
of groups that maximise a log-likelihood ratio merit function calculated using
the number of photons within and the total duration of the two candidate groups.
In other words, pairs of groups with the smallest intensity difference will be
combined. This merging of group pairs is repeated until all the groups have been
merged into a single group comprising all levels.

One of the limitations of AHC is that the outcome is highly dependent on the
initial state as well as outliers, and once two groups have been merged, the
decision is not challenged. For this reason, the outcome of the AHC --- the
combined levels corresponding to each number of possible groups --- is used as
an initial state for more advanced EM clustering, which serves to optimise the
intensity level groupings. In the EM algorithm, the log-likelihood that each
group would have a certain resulting intensity is calculated given the number of
photons in the group and the group's total duration as defined by the Poisson
probability density function. After the total probability is calculated, new
probabilities for each level relating to each group are calculated. 

Neither the AHC nor EM clustering method predicts the most likely number of
groups. The latter is done using the Bayesian information criterion
(BIC)~\cite{Watkins2005}. \textit{Full SMS} provides an interactive tool to
visualise the groups identified in each AHC--EM grouping scheme along with the
corresponding BIC value.

The screenshot in Fig.~\ref{fig:gui_grouping_example} is an example of
intensity-level grouping of data from an Alexa measurement. The top pane shows
the group intensity as dotted lines and the group boundaries as yellow lines. 
The resulting intensity ``bands" are indicated by alternating
light-blue and white bands. The BIC values for each grouping scheme are shown in
the bottom pane where the \textit{best} grouping scheme is indicated as a green
circle, representing the statistically most likely number of states, which
corresponds to the largest BIC value. The option that is currently applied is
indicated by the red outline, but the user can apply different solutions by
\textit{clicking} on any of the other circles. In this example, the best
solution is also applied. On the bottom right are options to include only the
ROI of each trace in the analysis, to group the levels of the current, selected
or all particles, and to ``apply" the groups to the levels (i.e., change each
level's intensity to that of its associated group). After performing the
grouping, the groups can also be viewed in the \textit{Intensity} tab
(Fig.~S6).

\begin{figure}[htp]
    \centering
    \includegraphics[height=3.8cm]{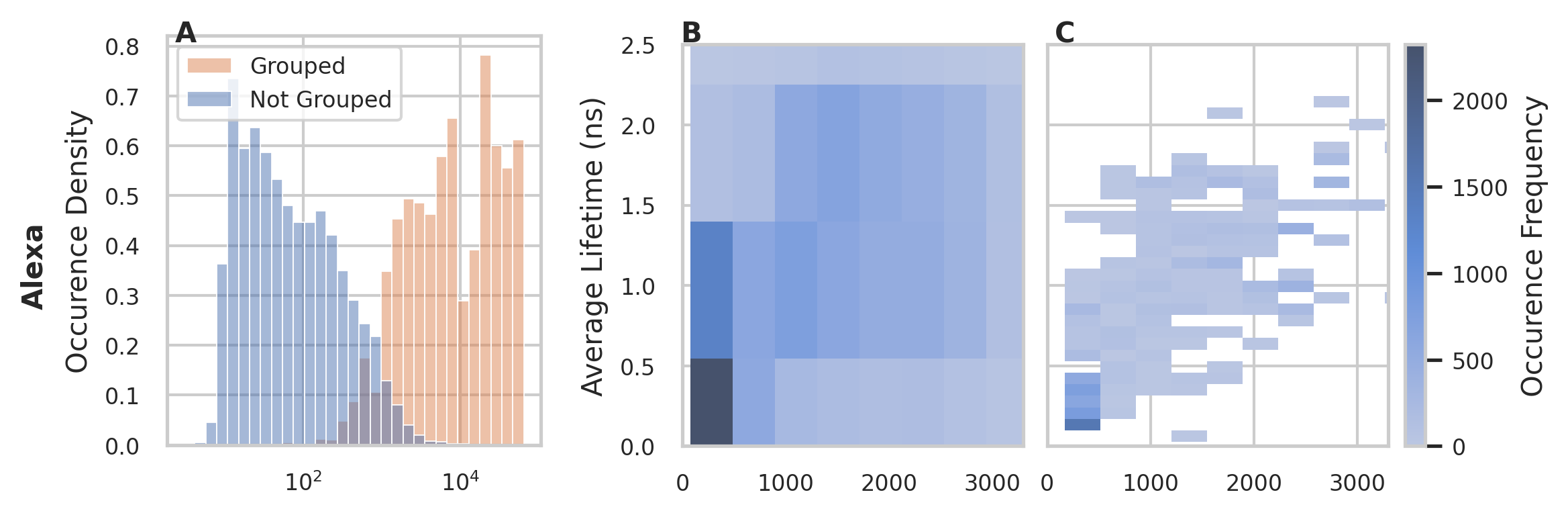}
    \includegraphics[height=3.8cm]{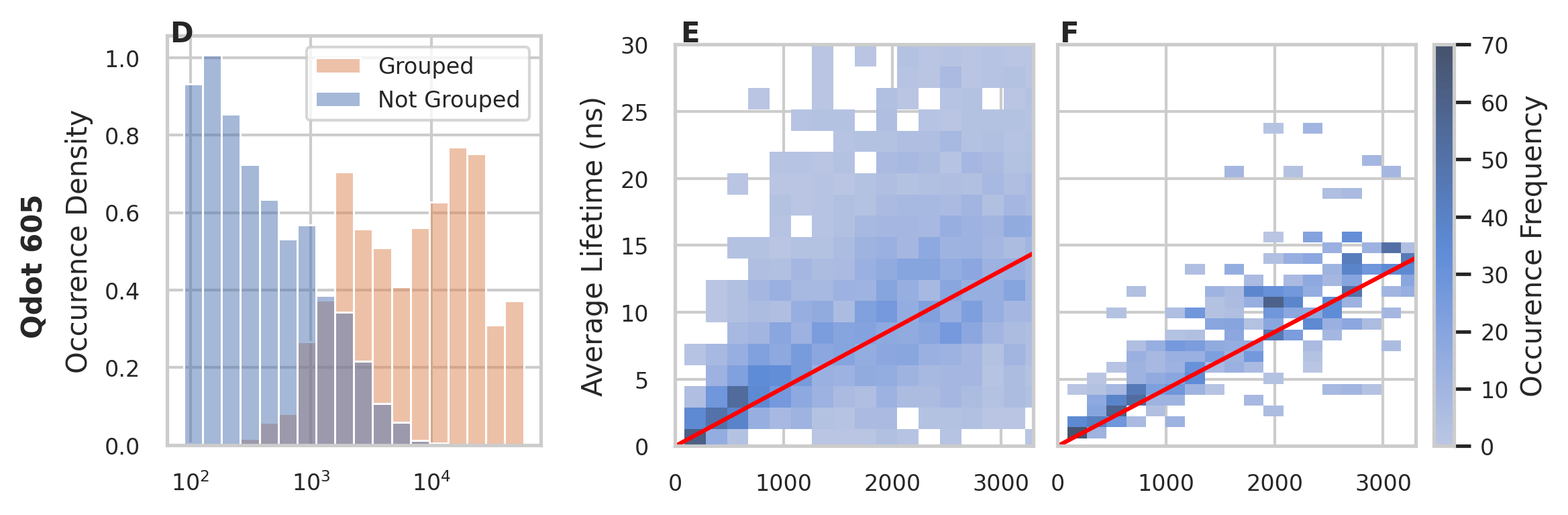}
    \includegraphics[height=3.8cm]{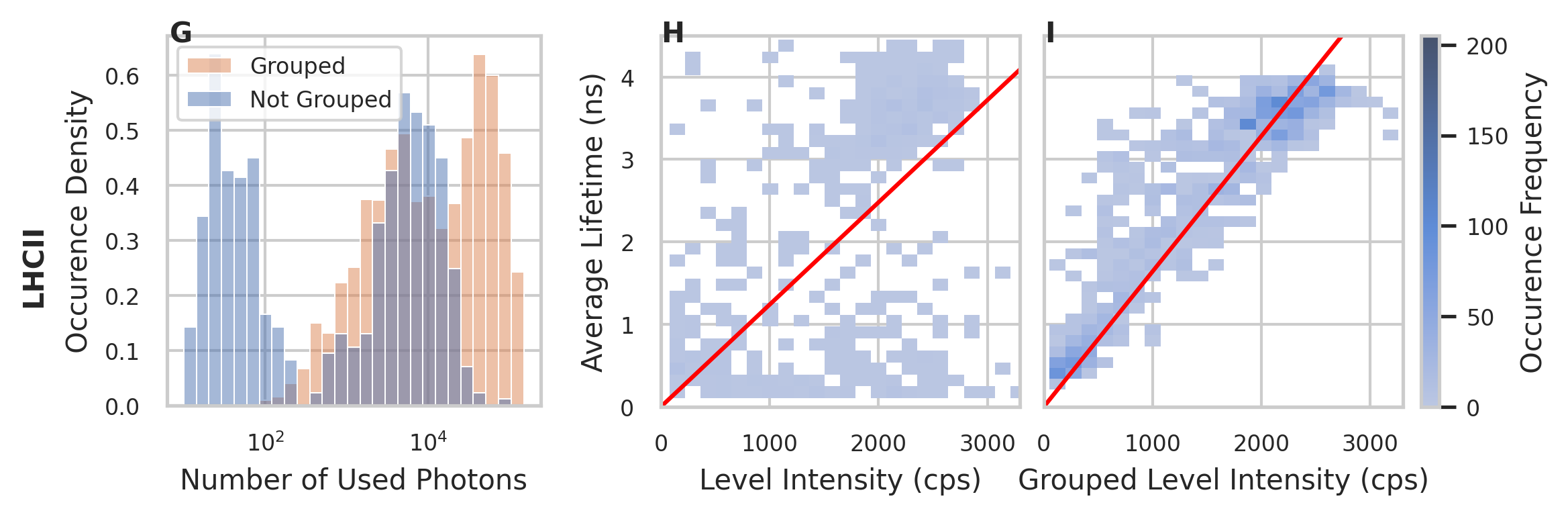}
    \caption{
        Lifetime fitting results of three datasets:
        Alexa Fluor 647 dye, Qdot 605 (CdSe/ZnS quantum dots), and light-harvesting complex
        II (LHCII). The first column shows the normalised distributions of the numbers of 
        photons used
        in each fitted level for both the non-grouped and grouped cases. The
        second and third columns depict the relationship between the weighted average of
        the fitted lifetimes for each level and the corresponding level intensity, where
        the second column is for the non-grouped levels and the third column is for the
        grouped levels. The red lines denote linear fits through the origin.
    }
    \label{fig:lifetime_fitting_examples}
\end{figure}

To demonstrate the advantage of grouping levels when fitting fluorescence
lifetimes, the fitting results of three datasets are compared, as shown in
Figure~\ref{fig:lifetime_fitting_examples}. The three samples for this case study
are Alexa (see above), Qdot 605, and light-harvesting complex II (LHCII). Qdot
605 is a commercially available type of CdSe/ZnS core-shell quantum dot and LHCII
is the main light-harvesting complex of plants, a pigment-protein complex
containing several energetically strongly connected chromophores that are
responsible for absorbing photons that provide the necessary energy to drive the
initial photochemical processes in the photosynthetic apparatus of higher
plants. The first column of these panels (A, D, and G) shows the distributions
of the numbers of photons used in each fitted level for both the non-grouped and
grouped cases. As the total number of levels between the non-grouped and grouped
cases differs, the Y-axis was normalised to facilitate comparison. The second
and third columns show two-dimensional distributions of the weighted average of
the fitted lifetimes and the level intensity for each level, with the second
column showing non-grouped levels and the third column showing grouped levels.
It is clear from panels A, D and G that the number of photons used for fitting
the lifetimes in the grouped levels is significantly larger than when no
grouping was done. The advantage of performing lifetime fitting on grouped
levels is particularly evident for the Qdot 605 and LHCII data, which shows the
expected predominantly linear relationships between lifetime and intensity
(panels F and I). This linear relationship is significantly broadened in the
Alexa data (panel C), which can be explained by the heterogeneous distribution
of the molecular orientations relative to the elliptically polarised excitation
light resulting in varying absorption cross sections and, consequently, altered
fluorescence intensities.

While this grouping method already offers several advantages, the resolved intensity 
levels of each trace are grouped in isolation from other traces, which could hamper 
the successful resolution of the underlying states. For example, not all states are likely 
accessed in each measurement. In \textit{Full SMS}, a global analysis 
can be done where each particle's ROI is appended into a single data set and then 
grouped, as shown in Figure~S7. Systematic intensity 
variations that may have resulted, for example, from focal drift during measurements 
will negatively affect the efficacy of a global grouping analysis. In a case where the 
lifetime and intensity are linearly correlated (like for Qdot 605 and LHCII in 
Fig.~\ref{fig:lifetime_fitting_examples}) the user has the option to perform an intensity 
normalisation as discussed in Section ~\ref{sec:data-filtering}, which will greatly 
improve the result of such an analysis. A sufficiently large dataset is likely to resolve 
true states but this operation is computationally expensive.

\subsubsection{\label{sec:second-order-correlation} Second-Order Correlation Function}

Single quantum emitters exhibit so-called antibunching of photons, as seen in
the second-order correlation function (\gtwo) of photon arrival
times~\cite{Mandel1995}. This is calculated as a cross-correlation between two
individual photon channels from detectors in a Hanbury Brown--Twiss
configuration since the dead time of a detector precludes the measurement of
coincident photons by a single detector. A perfect single emitter will never
emit two photons at the same time, giving a value of $g^{(2)}(0) = 0$. The number of 
independent emitters, $1/(1-g^{(2)}(0))$, is the reciprocal of the value of the 
``antibunching dip", $1-g^{(2)}(0)$. \textit{Full SMS} can calculate \gtwo\ for
two-channel data with an adjustable time window and bin size.

\begin{figure}[htp]
    \centering
    \includegraphics[width=1\linewidth]{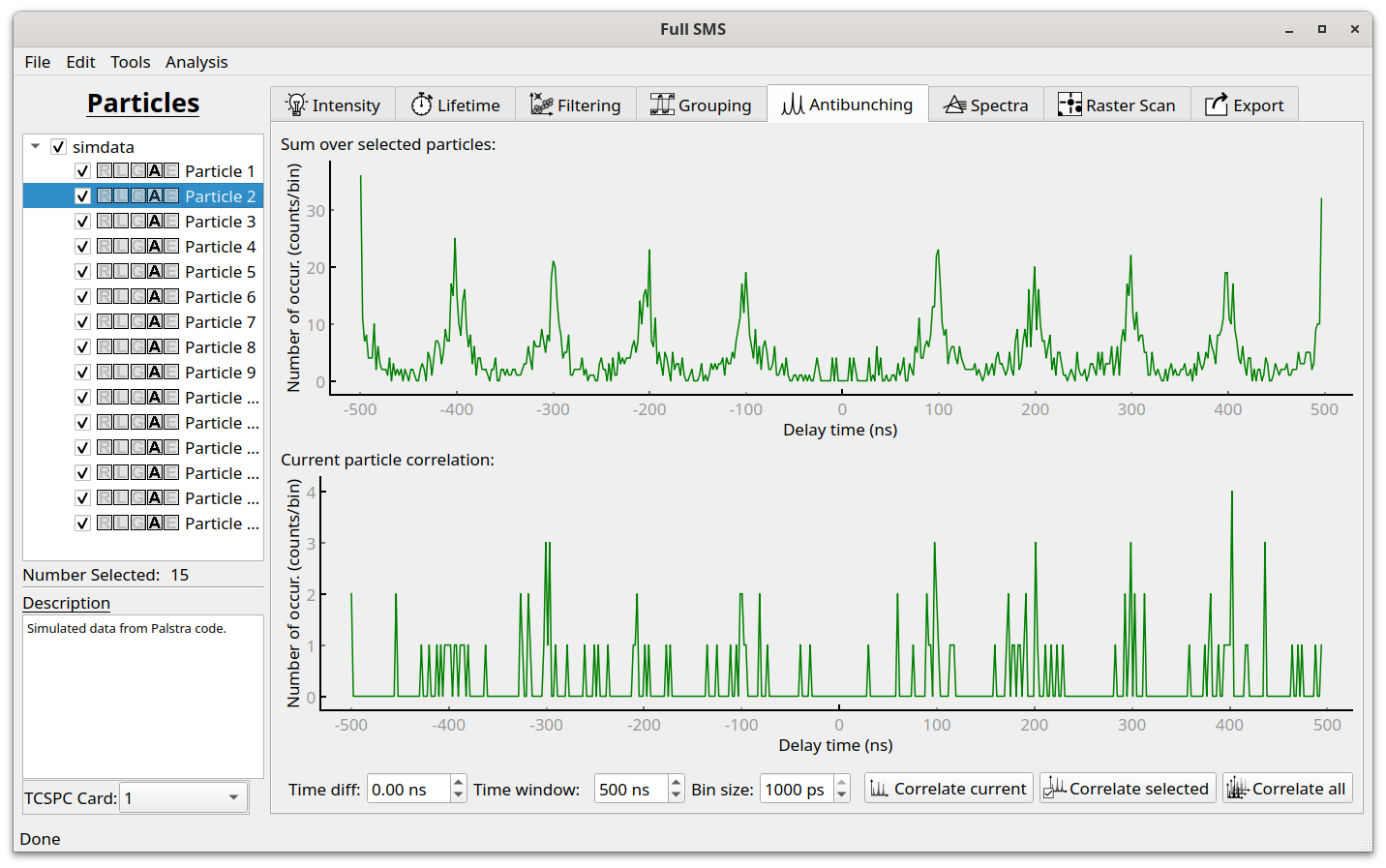}
    \caption{
        Antibunching analysis using second-order photon correlation histogram (\gtwo).
        The bottom pane shows the current particle's correlation histogram, while the 
        top pane shows the sum of correlations over all the selected particles, indicating 
        clear evidence of photon antibunching through the absence of a zero-delay peak.}
    \label{fig:abdemo}
\end{figure}

We demonstrate the analysis of photon antibunching using simulated data obtained
from the code of Palstra and Koenderink~\cite{Palstra2021} and converted to the
\textit{Full SMS} HDF5 format (Fig.~\ref{fig:abdemo}). In the
\textit{Antibunching} tab of the software, \gtwo\ can be calculated over an
adjustable time window and bin size. If there is a significant time delay between 
the two channels, this can be compensated for using the \textit{Time diff}
setting. The sum over all selected particles' \gtwo\ is shown in the top pane,
which is useful in the case of low-intensity or short measurements where
antibunching is not readily apparent in a single particle's correlation
histogram. Note that while the demonstration shows a simulation for pulsed
excitation, the calculation of \gtwo\ following continuous-wave excitation is
identical and can, therefore, also be done by \textit{Full SMS}.

\subsubsection{\label{sec:spectra} Display of Spectra}

\begin{figure}[htp]
    \centering
    \includegraphics[width=1\linewidth]{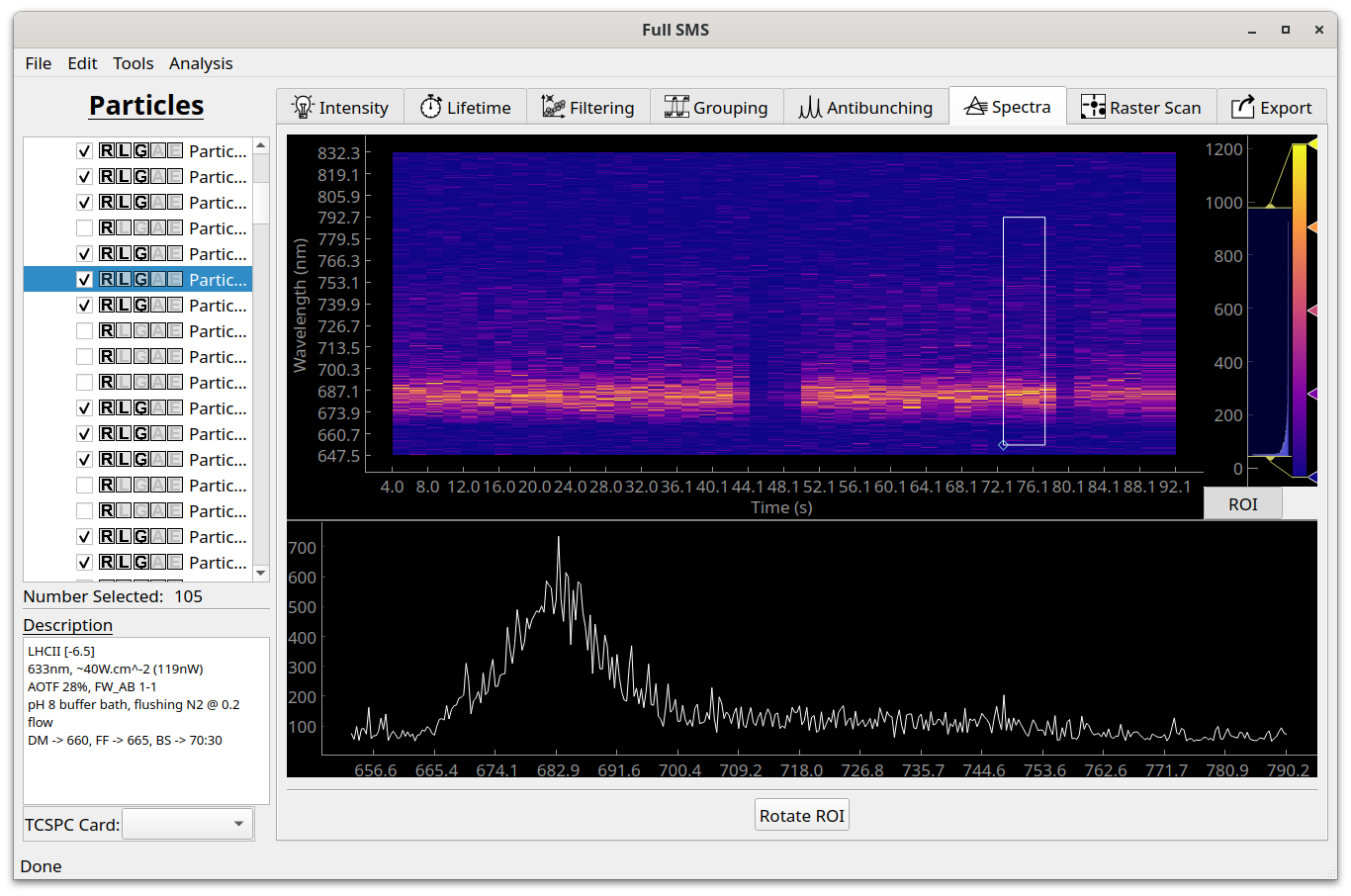}
    \caption{
        Example of a spectral measurement of a single LHCII pigment-protein complex, 
        showing a full spectral time trace (top) and temporally summed spectrum of 
        the selected ROI (bottom).
    }
    \label{fig:gui_spectra}
\end{figure}

In the \textit{Spectra} tab of the software (Fig.~\ref{fig:gui_spectra}),
spectral time traces can be viewed, which are displayed as a 2D colour map of
the photon counts (intensity) as a function of the bin time and wavelength on
the $X$- and $Y$-axes, respectively. The colour scale can be adjusted for
optimal contrast by setting the upper and lower thresholds as well as the
relative intensities of set colours along the scale. A histogram of pixel
intensities is shown to assist in this regard. An ROI can be defined, allowing a
plot of intensity as a function of time or wavelength to be displayed.

\subsubsection{\label{sec:raster-scan} Display of Raster Scan Images}

The \textit{Raster Scan} tab of the software
(Fig.~S8) displays raster-scan images. In SMS
experiments, these images are mostly used to locate particles to be measured.
Therefore, the functionality of this tab is currently limited to displaying
images and indicating the position of the currently selected particle. The
images can, however, easily be exported if further image analysis or processing
is required.

\subsubsection{\label{sec:export} Data Export}

\begin{figure}[htp]
    \centering
    \includegraphics[width=1\linewidth]{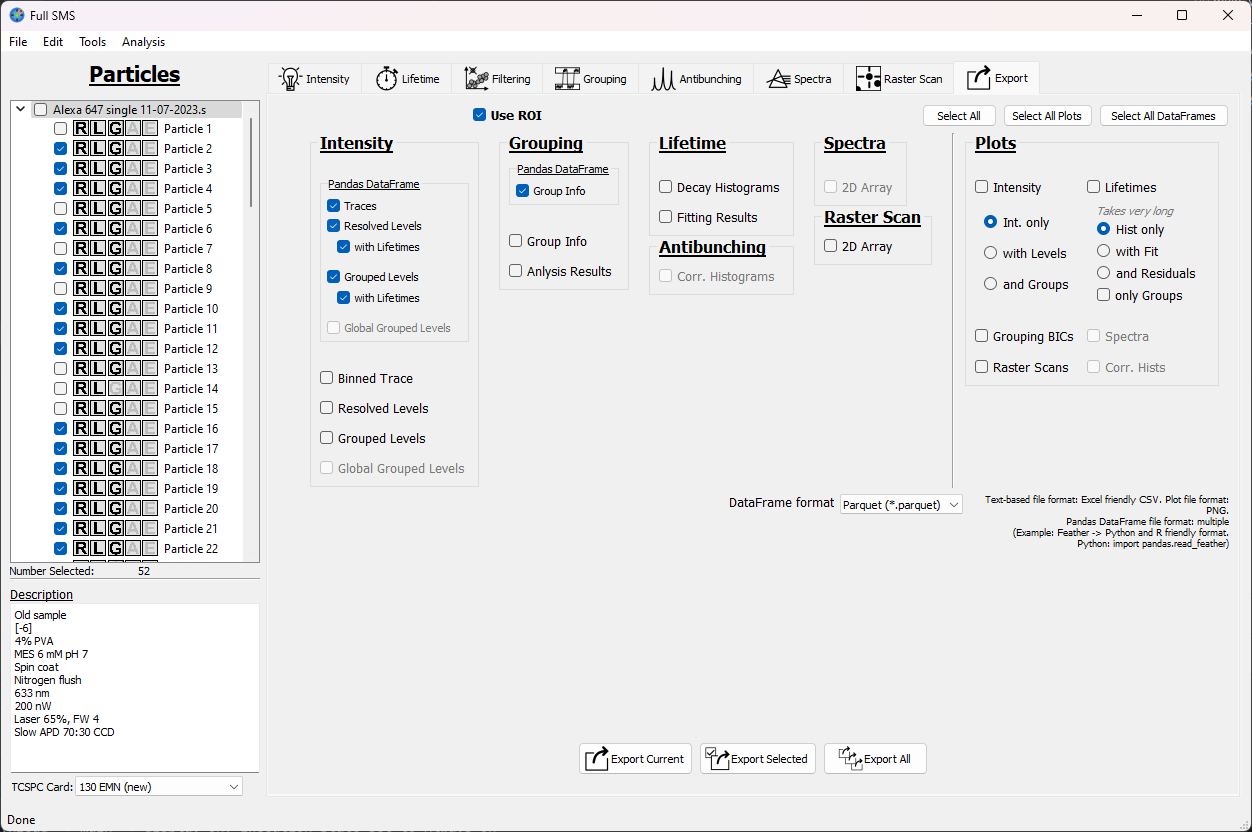}
    \caption{Exporting tab of the software. The needed data can be selected, which can then be 
    exported for the current, selected, or all particles.}
    \label{fig:gui_exporting}
\end{figure}

In the \textit{Export} tab of the software (Fig.~\ref{fig:gui_exporting}),
the analysed data can be exported for further analysis. Binned intensity traces,
resolved and grouped levels, and lifetime fitting results, can all be exported
as plain text, or in the form of a parquet file (a commonly used open-source column-oriented 
binary data storage format maintained by Apache), which enables more
convenient downstream processing for users familiar with Pandas, a Python
library for data manipulation and analysis, which can easily read and load parquet files. 
Apart from this final export, it is also possible to save the current state of 
the analysis in order to continue working on it at a later time. This includes 
the binned intensity, resolved, grouped and fitted levels, as well as the current 
particle selection.

\subsection{\label{sec:tools} Additional Tools}

In addition to the previously mentioned advantages of \textit{Full SMS}, some
additional \textit{tools} have been developed. These tools add significant
functionality to the analysis or enhance the user experience by automating
common tasks.

\subsubsection{\label{sec:data-conversion} Measurement Data Conversion}

The native format of measurement data for which this software suite has been
designed (HDF5) is custom in that the particular structure and naming used were
arbitrary choices that needed to be made. As mentioned above, \textit{Full SMS}
uses a custom HDF5 file format --- custom in the sense that the specific grouping
and naming of data inside the hierarchical format are customised. Therefore, if
measurements are made in a format that differs from this, either in file format
or structure, it will be necessary to convert the original data into the format
that \textit{Full SMS} can read. This can be done \textit{ad hoc} or built into
the software. As an example of the latter, we included a conversion tool that
converts the measurements made on a different experimental setup (in PicoQuant's
.pt3 format) to the format needed for this software. This also acts as an
example for future contributors of how such a conversion can be done. Our
software can also easily be extended to allow the analysis of Photon-HDF5 data.

\subsubsection{\label{sec:trace-roi} Trace Region of Interest (ROI)}

When performing SMS measurements, depending on the nature of the sample being
studied, it is almost certain that a single particle cannot be measured for an
indefinitely long time. This is usually due to photodamage, as the excitation
intensities are typically large. Particles often bleach during a measurement
run, and therefore it is often valuable to select a subsection of the
measurement for analysis, excluding the data originating from a bleached
particle. However, in cases where datasets contain many measurements, applying
some user-defined criteria to determine the ROI for each
measurement individually is often a tedious task. To aid this task, a tool has
been included to automate the application of an ROI over the
entire dataset (Fig.~S9).

The user defines a lower intensity threshold and a lower duration threshold. 
Resolved levels are considered until the last one that satisfies both thresholds. 
Each data set is analysed from back to front, and the point (if any) at which both of these
thresholds are broken is set to be the end of the ROI. An example
is shown in Fig.~\ref{fig:gui_resolved_trace}, where the ROI is shown by the
green shaded area, excluding the presumably bleached state at the end of the trace.

\subsubsection{\label{sec:photon-burst-detection} Photon Burst Detection}

In certain types of SMS measurements, very short-lived spikes in intensity are
recorded, a phenomenon that may be caused by cosmic rays or unwanted fluorescent
particles diffusing through the focal volume. In most cases, such spikes skew
the analysis and therefore are removed. \textit{Full SMS} provides the user with
two methods of identifying and removing photon bursts
(Fig.~S10). The first is based on the standard 
deviation ($\sigma$) of the mean intensity of each resolved level. A multiple 
of $\sigma$, as chosen by the user, above the mean intensity is used as a threshold 
definition. Alternatively, the user can define a manual intensity threshold. In 
either case, if any levels exceed the threshold, they are identified as potential 
photon bursts and can be removed.

\section{\label{sec:application-architecture} Application Architecture}

\subsection{\label{sec:general-architecture} General Architecture}

The program was written completely in Python, using a package called PyQt5 to
interface the Qt GUI framework. The Qt GUI framework was chosen as it has rich
features, works across all platforms (Windows/Mac/Linux) and allows rapid
development of GUI windows using a GUI builder (Qt Designer). However, the
primary complexity of adding a GUI is not in the design of the windows but in
the underlying framework that supports it.

\textit{Full SMS} was designed using object-oriented programming and makes use
of a \textit{controller} architecture, in which several class instances are
responsible for separate aspects of the program. The interaction with the GUI,
as well as the functional implementations of the analysis, are handled by
several different controller classes. For example, the intensity controller is
responsible for converting time series to binned intensity traces and
interacting with the GUI to display the selected intensity trace. However, much
of the advanced functionality that is used within the controllers is implemented
in separate modules. For example, the CPA code is found within
its own module but is called from within the intensity controller. In this
manner, the code is organised into logical sections.

\subsection{Multi-processing}

One of the critical limitations of the Python language\footnote{This limitation
is still the case as of Python version 3.12, but is planned to be removed in
some future release.}, as implemented by CPython\footnote{CPython is the most
common implementation of Python written in C.}, is that it makes use of a global
interpreter lock (GIL). The GIL is a necessary restriction that allows only
one thread to execute code for a single Python process, effectively limiting
the CPU core usage to a single core. Multi-threading is still possible by allowing
a single process to switch between threads rapidly. In the case of a
GUI, this allows the interface to remain responsive to the user, even if a
separate thread is running to, e.g., perform an analysis. However, the GIL
prevents effective usage of modern computing power when performing analyses that
allow for parallelisation and would benefit from multiple CPU cores.

Several methods have been developed to achieve multi-core processing, which, in
most cases, involve starting multiple Python processes. The primary complexity
is then the communication between processes.

The approach that the \textit{Full SMS} software uses is as follows. Once the
user interacts with the GUI, a thread is started to perform the analysis while
allowing the GUI to remain reactive. The analysis thread then uses Python's
built-in \verb|multiprocessing| module to start several Python processes, which
will here be referred to as worker processes. The number of worker processes
created is automatically scaled to the number of CPU cores the system has
available. The use of queues achieves communication between processes. Queues
are populated by the analysis thread with pairings of copies of a function with
serialised versions of the data the function should be applied to. The functions
and serialised data are loaded from the queue by the worker processes and
executed. Afterwards, the results are added to another queue in the form of
serialised objects. The unfortunate consequence is that the parent process must
consolidate the returned results into the original data structures. This can be
done easily for simple objects, but for complex objects, as is the case for most
of the objects in this software, great care needs to be taken to integrate all
the relevant parts of the returned object. On completion of all the parts of the
analysis, the parent process terminates the worker processes and the thread
finishes.

\subsection{Distribution and Compatibility}

The repository for \textit{Full SMS} is publicly hosted on GitHub and can be
accessed at \url{http://github.com/BioPhysicsUP/Full_SMS}. The easiest way to
run the software is via a Windows installer that can be downloaded from the
repository releases page. This installer does not have any prerequisites but
is currently only available for Windows. Instructions for running from source
can be found in the GitHub README. Detailed documentation of the software,
including how to contribute, is available at
\url{https://up-biophysics-sms.readthedocs.io/en/latest/index.html}.

\section{\label{sec:conclusion} Conclusions}

We have described a new GUI-based application for advanced analysis of
single-molecule spectroscopic data. It uses a custom HDF5 file format
that is uniquely suited to multi-parameter measurements on single particles. The
software is user-friendly, requiring no programming knowledge to use, though
also being fully open source and easily extendable. It allows the analysis of
photon-by-photon data to extract fluorescence intensity change points and groups
the resulting intensity levels using agglomerative hierarchical clustering. It
also allows fitting fluorescence lifetimes of individual and grouped levels, the 
calculation of the second-order intensity correlation function, data filtering 
options, and a few additional features.

\textit{Full SMS} is highly suited to further extensions. An extension of the
multi-channel functionality to include multi-channel change-point
detection~\cite{Wilson2021} would be useful for analysing smFRET data. The
lifetime functionality could be extended to allow fitting with more than three
exponential components and with decay models other than a multi-exponential,
such as lifetime distributions. Functionality could be added to analyse spectra
and for analysing fluorescence correlation spectroscopy (FCS) and real-time
feedback-driven single-particle tracking~\cite{vanHeerden2022} data. Adding the
ability to interface directly with other data formats, such as
Photon-HDF5~\cite{Ingargiola2016}, would enable more groups to start using the
software immediately.

\section{\label{sec:method} Experimental Methods}

\subsection{Experimental setup}

As described elsewhere \cite{Kyeyune2019}, a pulsed supercontinuum laser
(Fianium, SC400–4–PP) with a repetition rate of 40 MHz was used. The desired
central wavelength of excitation depended on the sample (see below) and was
determined by an acousto-optic tunable filter (AOTF) (Crystal Technology, Inc.).
A combination of a linear polariser (LPVISB050-MP, Thorlabs) and a quarter-wave
plate ($\lambda$/4 485-630, Achromatic Retarder, Edmund Optics) produced
near-circularly polarised laser pulses that were subsequently passed through a
spatial filter. The beam was reflected by a dichroic mirror (see below) and
focussed into a near-diffraction-limited by a 1.45 numerical aperture (NA)
oil-immersion objective (Plan-Fluor Apo $\lambda$ 100$\times$, Nikon). Samples
were mounted on a three-axis piezo nanopositioning stage (LPS200, Mad City
Labs). The fluorescence was collected by the same objective and focussed onto a
100-$\mu$m pinhole to filter out light not originating from within the focal
plane. An appropriate fluorescence filter was used to improve the
signal-to-background ratio further depending on the fluorescence wavelength of
the sample (see below). The emitted light was split into two beams, 70\% of
which was reflected and focussed onto a single–photon avalanche photodiode
(PD–050–CTE, Micro Photon Devices PDM, IRF $\sim$ 128 ps) that generated a
series of electronic pulses, which in turn was measured by a time-correlated
single-photon counting (TCSPC) module (SPC-130-EM, Becker \& Hickl). The
remaining 30\% was transmitted to measure the fluorescence spectra by dispersing
the light using a grating (GR25–0608, 600/mm, 750~nm blaze, Thorlabs) and
thereafter focussing the light onto an electron-multiplying charge-coupled
device camera (EMCCD) (iXon$_3$, Andor) with an integration time of one second.

\subsection{Samples}

Alexa Fluor 647 carboxylic acid (Thermo Fisher Scientific) was diluted to ${\sim}
25$~pM in 6~mM 2-(N-morpholino)ethanesulfonic acid (MES) buffer (pH 7)
containing 4\% (w/w) poly(vinyl alcohol), and spin-coated onto a glass coverslip. The
excitation power and wavelength were 200~nW and 633~nm, respectively. The
dichroic mirror and fluorescence filter were FF650-Di01-25x36 (Semrock) and
FELH0650 (Thorlabs), respectively.

Qdot 605 (carboxylic acid conjugate) (Thermo Fisher Scientific) was diluted to
${\sim}80$~pM in 10~mM MES buffer (pH 7) containing 0.2 mM~MgCl$_2$ and 0.05\%
(w/v) Tween-20. A small droplet was applied to a coverslip treated with
poly-L-lysine (PLL) and another coverslip was placed on top. The excitation
power and wavelength were 140 nW and 488 nm, respectively. The dichroic mirror
and fluorescence filter were 605dcxt (Chroma Technology Corp.) and 600LPF (Edmund Optics), respectively.

LHCII from spinach, isolated using the protocol in Ref.~\cite{Xu2015}, was
diluted to ${\sim} 3$~pM in 20 mM 4-(2-hydroxyethyl)-1-piperazineethanesulfonic
acid (HEPES) buffer (pH 8) containing 0.03\% (w/v) n-dodecyl-α-D-maltoside
(α-DM) and 1 mM MgCl$_2$. A small droplet was applied to a coverslip treated
with PLL, and the coverslip was placed in a custom sample chamberfilled with an
additional buffer-detergent-salt solution. To remove oxygen and prolong
measurement times, a glucose/glucose oxidase/catalase mixture was used along
with flushing of the sample chamber with N$_2$ gas. The excitation power and
wavelength were 119 nW and 633 nm, respectively. The dichroic mirror and
fluorescence filter were TX660 and ET665lp, respectively (both from Chroma Technology Corp.).

\backmatter

\bmhead{Supplementary Information}
The Supplementary Information contains additional GUI screenshot examples.

\bmhead{Acknowledgments}

We thank the UP Biophysics group members for testing the software and suggesting 
improvements. We gratefully acknowledge Michal Gwizdala for the LHCII isolation. 

\section*{Declarations}

\bmhead{Funding}

JLB was supported by the Vrije Universiteit Amsterdam–NRF Desmond Tutu Programme. 
BvH was supported by the National Research Foundation (NRF), South Africa (grant 
nos. 115463, 120387), the South African Academy for Science and Art, NITheCS, and 
the Fulbright Programme. TPJK acknowledges funding from the NRF (grant nos. 87990, 
94107, 109302, 110983, 112085, 120387, and 137973), the Photonics Initiative of 
South Africa, the Rental Pool Programme of the Council for Scientific and Industrial 
Research's Photonics Centre, South Africa, and the University of Pretoria's Research 
Development Programme, Strategic Research Funding, and Institutional Research Theme 
on Energy.

\bmhead{Competing interests} 

The authors declare no competing interests.

\bmhead{Availability of data}

Data is available from the authors upon request. 

\bmhead{Code availability}

Code is available at \url{http://github.com/BioPhysicsUP/Full_SMS}.

\bmhead{Authors' contributions}

Joshua L. Botha and Bertus van Heerden contributed equally to this work. All
authors conceived the study. Joshua L. Botha and Bertus van Heerden wrote the
software, performed experiments and analysed data. Tjaart P.J. Kr\"uger acquired
funding, supervised research, and administered the project. Joshua L. Botha and
Bertus van Heerden wrote the original draft. Bertus van Heerden and Tjaart P.J.
Kr\"uger reviewed and edited the manuscript with input from Joshua L. Botha. All
authors read and approved the final manuscript.

\bibliography{manual_references} 

\appendix

\section{Supplementary Information}
\subsection{Graphical User Interface Examples}

\begin{suppfigure}[h]
    \centering
    \includegraphics[width=0.95\linewidth]{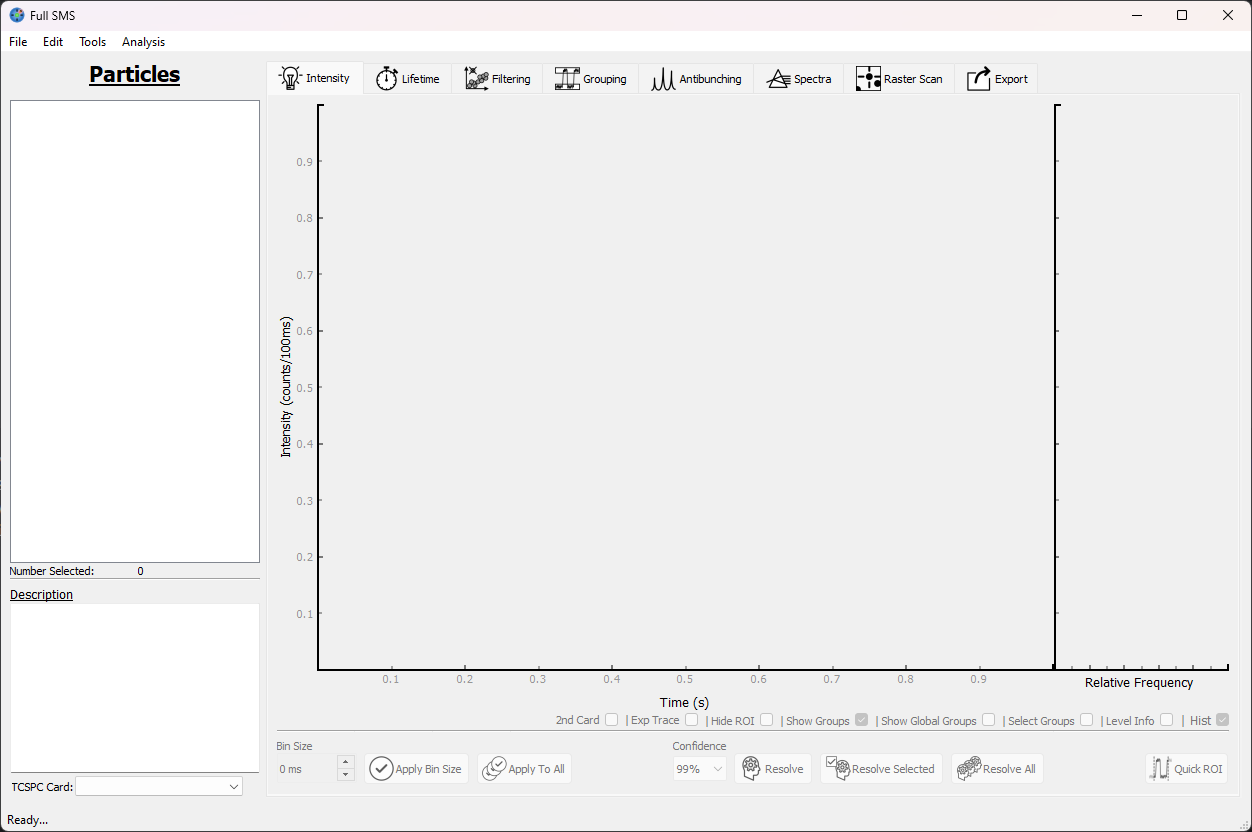}
    \vspace*{2mm}
    \caption{
        Main window of \textit{Full SMS} without any data file loaded.
        The analysis is separated into separate tabs for each major operation, starting 
        with the \textit{Intensity} tab and ending with the \textit{Export} tab.
    }
    \label{suppfig:gui_main_window}
\end{suppfigure}

\begin{suppfigure}[h]
    \centering
    \includegraphics[width=0.95\linewidth]{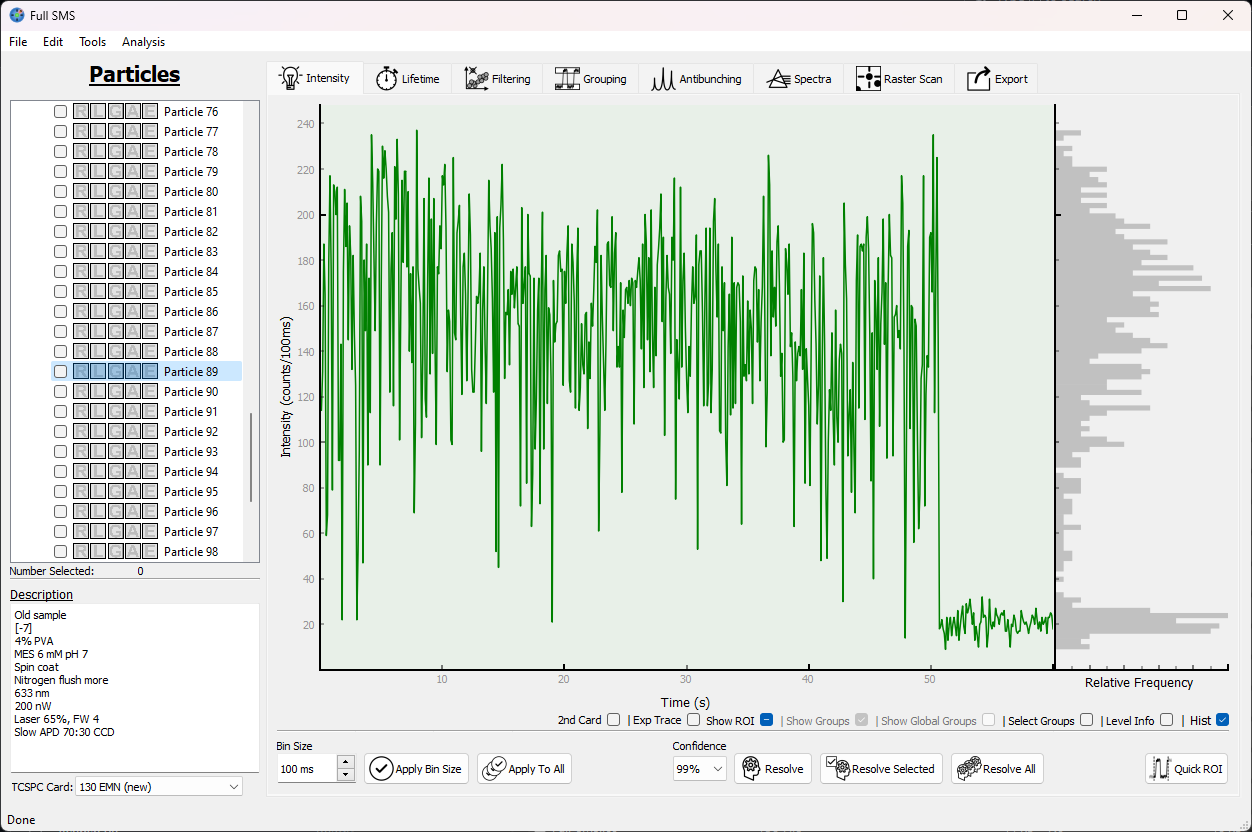}
    \vspace*{2mm}
    \caption{Intensity trace example of Alexa as shown in the \textit{Intensity} tab, 
    showing a binned intensity trace before the resolving of the intensity levels is 
    actioned. The same trace is shown in Fig.~1 with the 
    levels resolved.}
    \label{suppfig:gui_trace_example}
\end{suppfigure}

\begin{suppfigure}[h]
    \centering
    \includegraphics[width=0.95\linewidth]{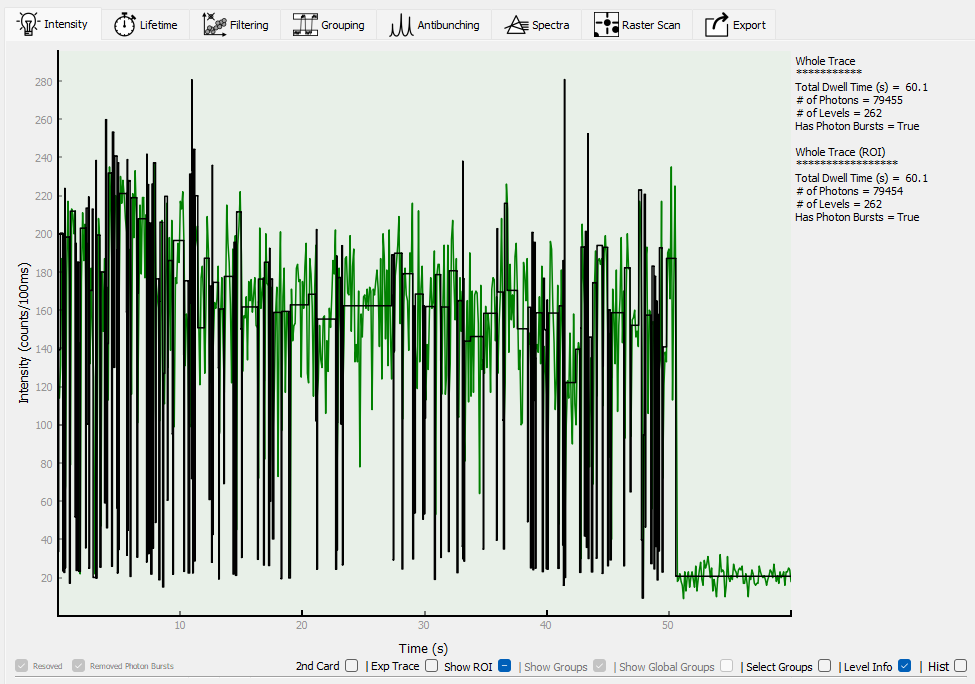}
    \vspace*{2mm}
    \caption{Level information (selected via \textit{Level Info}) of the Alexa intensity 
    trace shown in Fig.~1, showing the total level dwell time, 
    number of photons, and number of levels.}
    \label{suppfig:gui_level_info_whole_trace}
\end{suppfigure}

\begin{suppfigure}[h]
    \centering
    \includegraphics[width=0.95\linewidth]{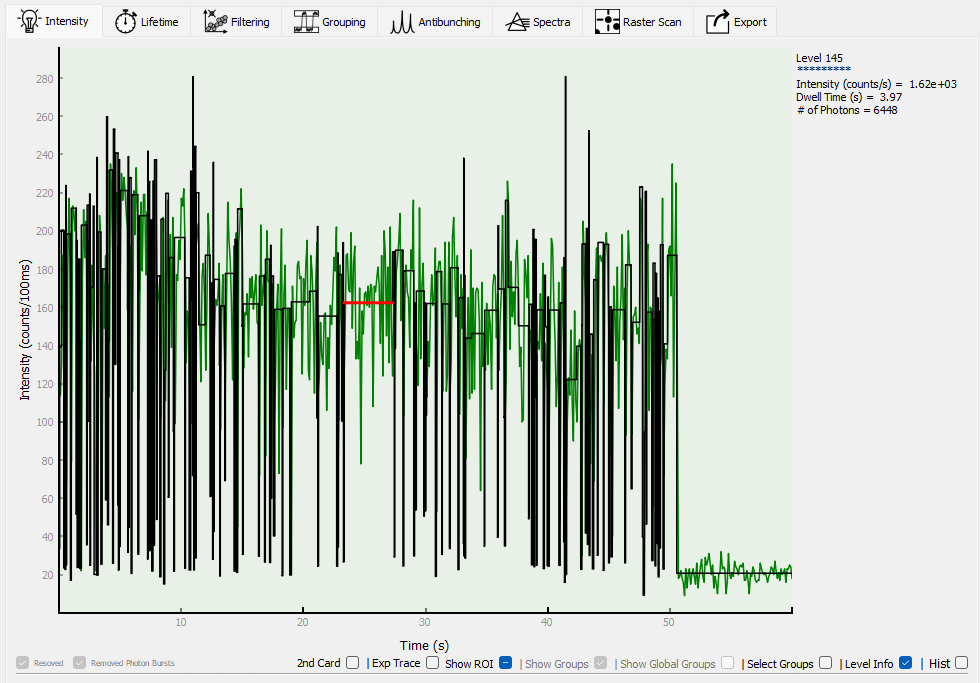}
    \vspace*{2mm}
    \caption{Level information (selected via \textit{Level Info}) of the Alexa intensity 
    trace shown in Fig.~1, showing the level intensity, dwell 
    time and number of photons.}
    \label{suppfig:gui_level_info_single_level}
\end{suppfigure}

\begin{suppfigure}[h]
    \centering
    \includegraphics[width=0.95\linewidth]{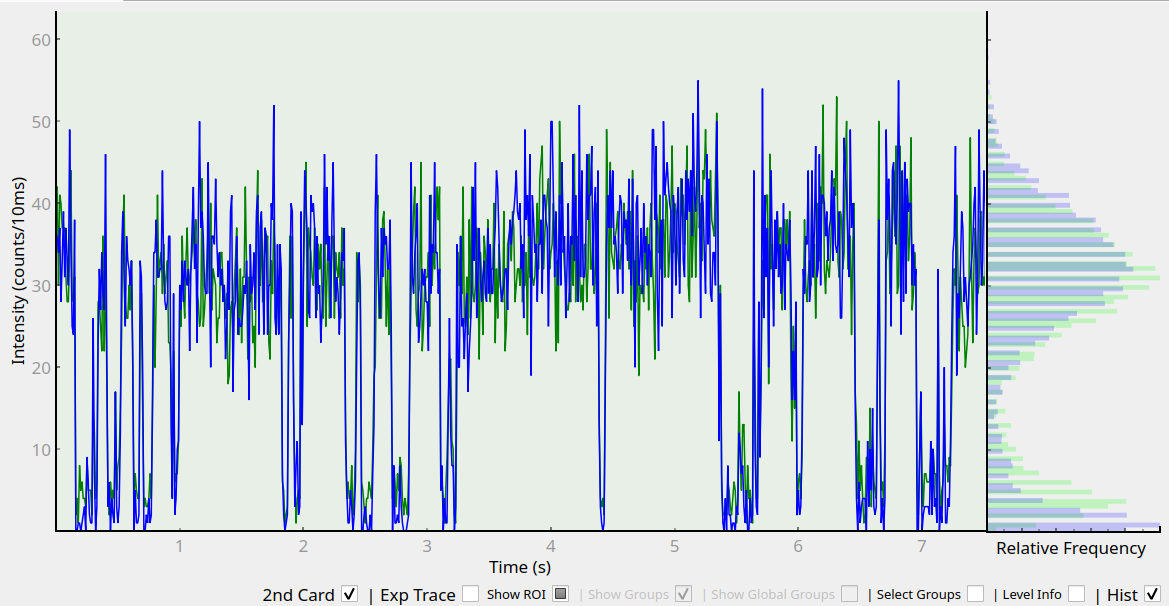}
    \vspace*{2mm}
    \caption{Two-channel intensity trace example, showing the binned intensity 
    traces from each channel as a separate colour.}
    \label{suppfig:2channel}
\end{suppfigure}

\begin{suppfigure}[h]
    \centering
    \includegraphics[width=0.95\linewidth]{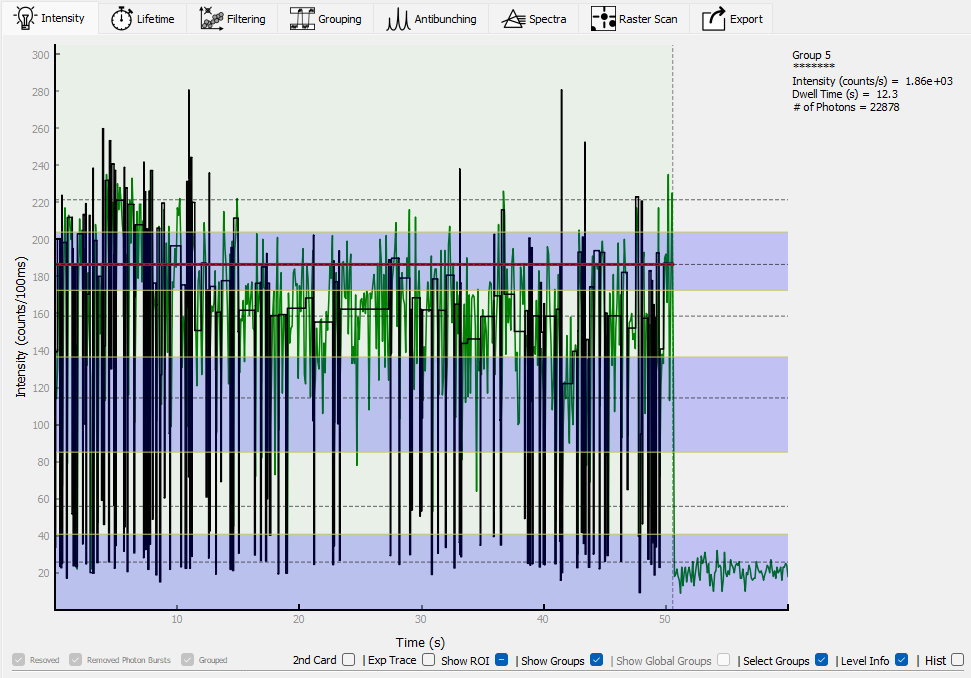}
    \vspace*{2mm}
    \caption{Group selection example using the groups shown in Fig.~5. 
    The group's intensity, dwell time, and number of photons are displayed.}
    \label{suppfig:gui_group_selected}
\end{suppfigure}

\begin{suppfigure}[h]
    \centering
    \includegraphics[width=0.95\linewidth]{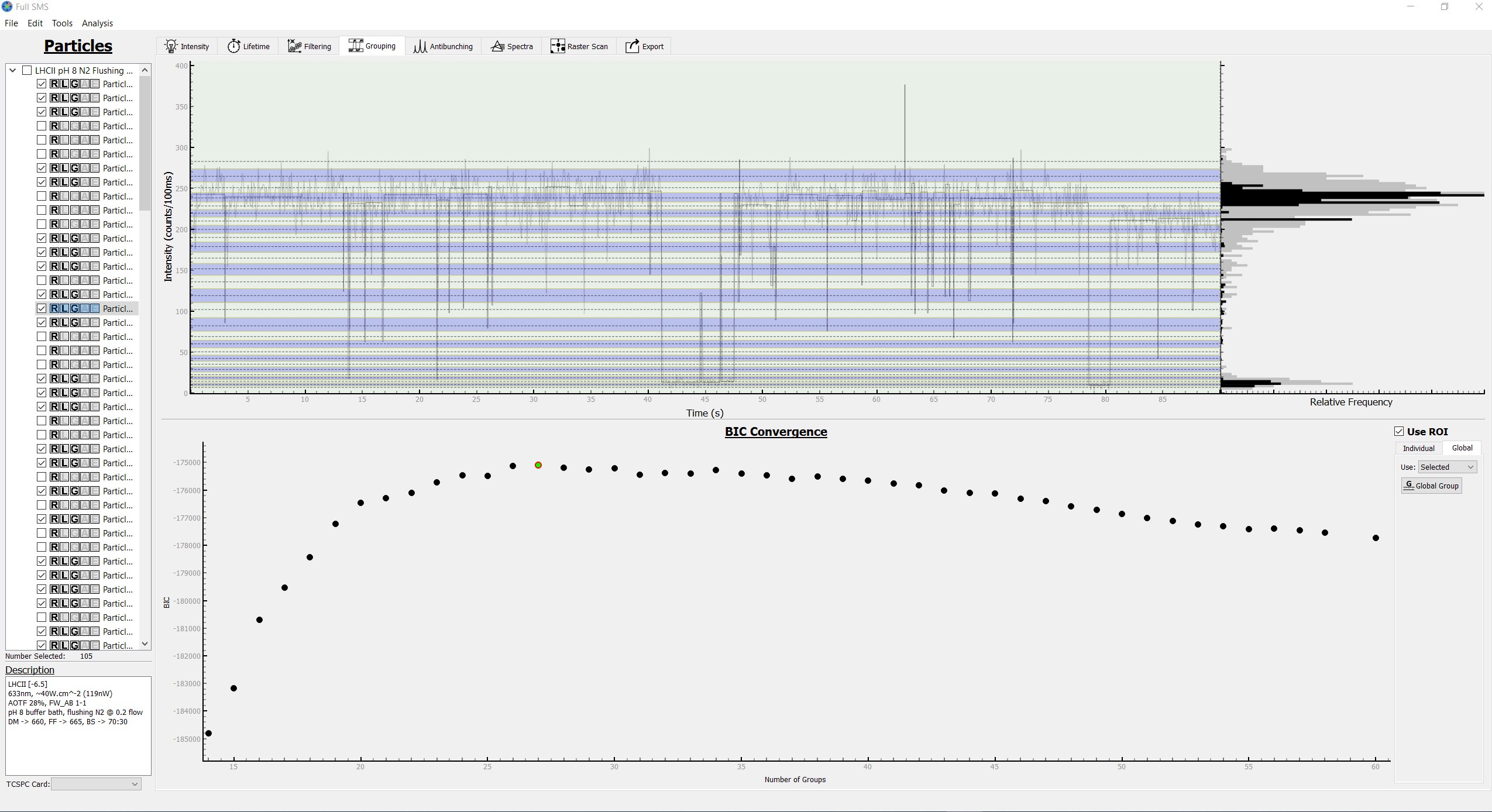}
    \vspace*{2mm}
    \caption{Global grouping example of the data of 105 LHCII complexes. The bottom pane 
    shows the progression of the grouping of the appended data, which originally comprised 
    $\sim 10^{5}$ individual intensity levels. The largest BIC value corresponds to 27 
    distinct states (down from $\sim 10^{5}$ states), some of which are not accessed in the 
    intensity trace in the upper pane.}
    \label{suppfig:gui_global_grouping}
\end{suppfigure}

\begin{suppfigure}[h]
    \centering
    \includegraphics[width=0.95\linewidth]{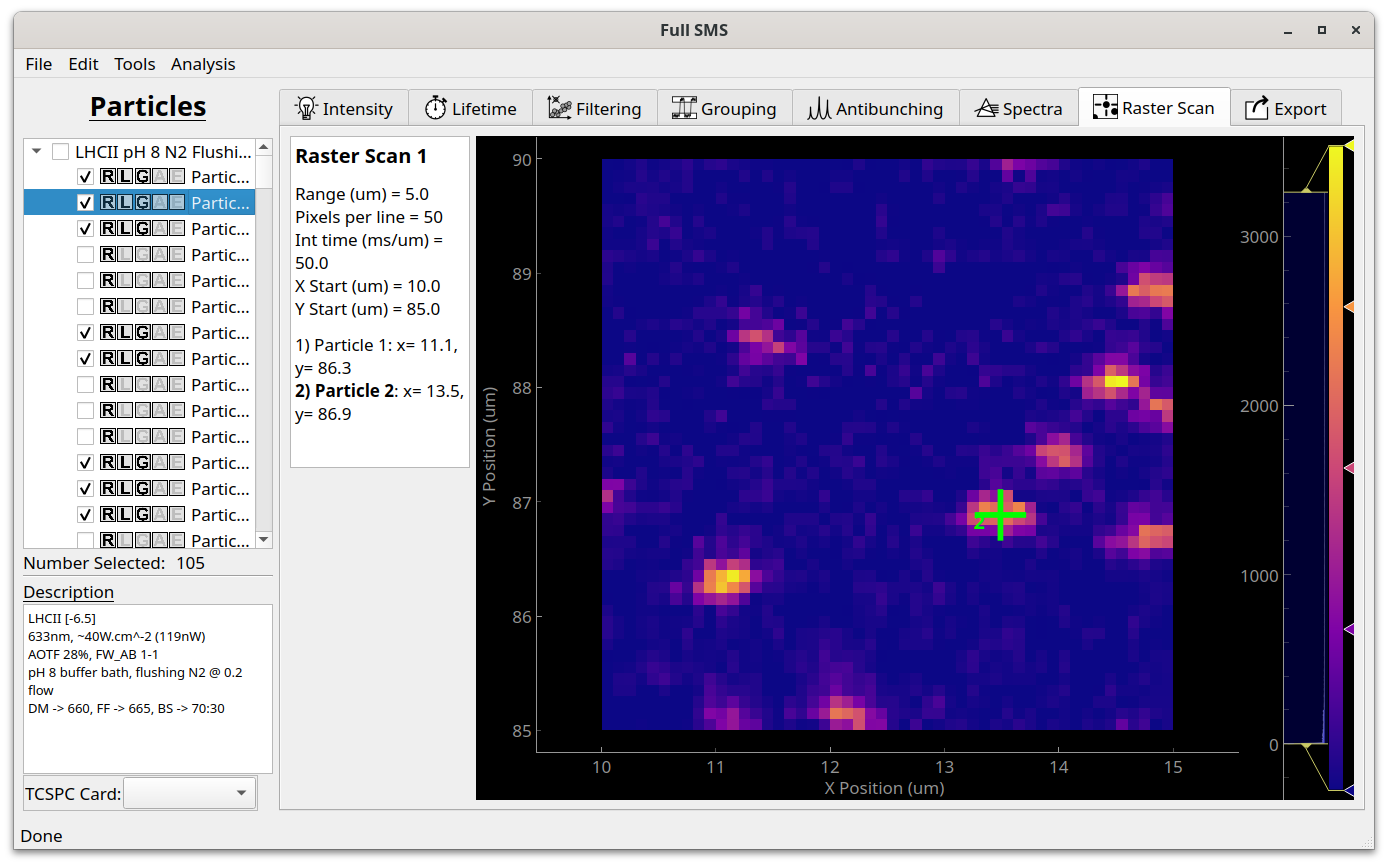}
    \vspace*{2mm}
    \caption{Example of a raster-scan image of a few LHCII complexes immobilised on a glass 
    coverslip via poly-L-lysine, indicating the position of the currently selected particle 
    with a green plus. The raster-scan details and coordinates of selected particles are 
    indicated on the left of the image.}
    \label{suppfig:gui_raster_scan}
\end{suppfigure}

\begin{suppfigure}[h]
    \centering
    \includegraphics[width=0.95\linewidth]{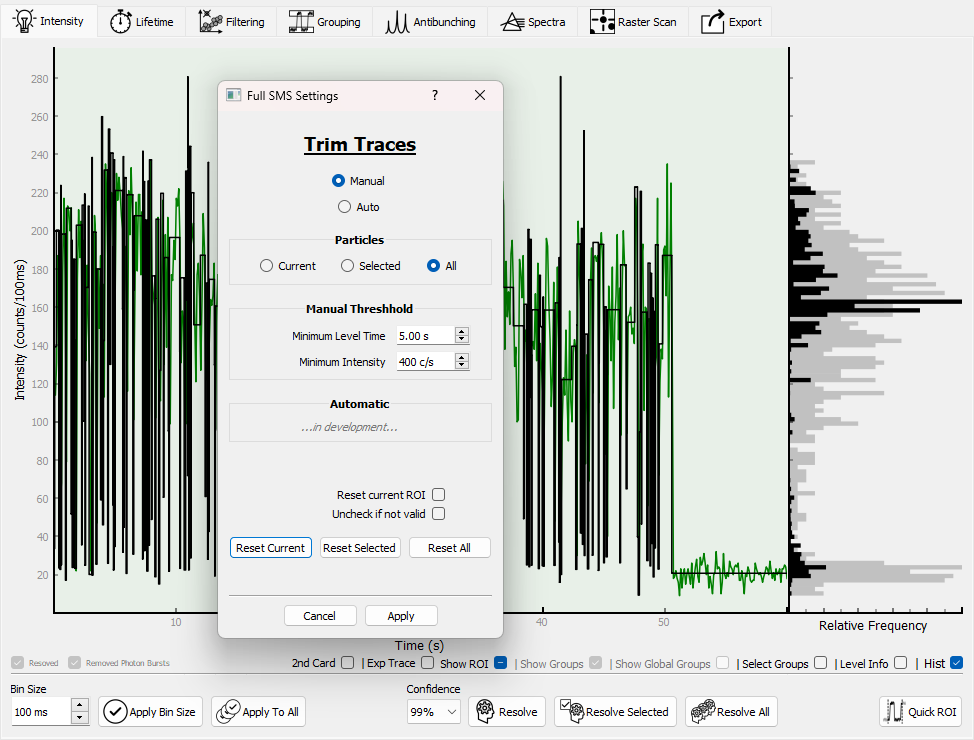}
    \vspace*{2mm}
    \caption{Trace trimming tool dialog with the Alexa intensity trace shown in 
    Fig.~1. The end of a trace is trimmed if it is below the given 
    minimum intensity and longer than the given minimum dwell time. An automated choice of 
    these parameters is a future extension that is currently under development.}
    \label{suppfig:gui_trim_trace_dialog}
\end{suppfigure}

\begin{suppfigure}[h]
    \centering
    \includegraphics[width=0.95\linewidth]{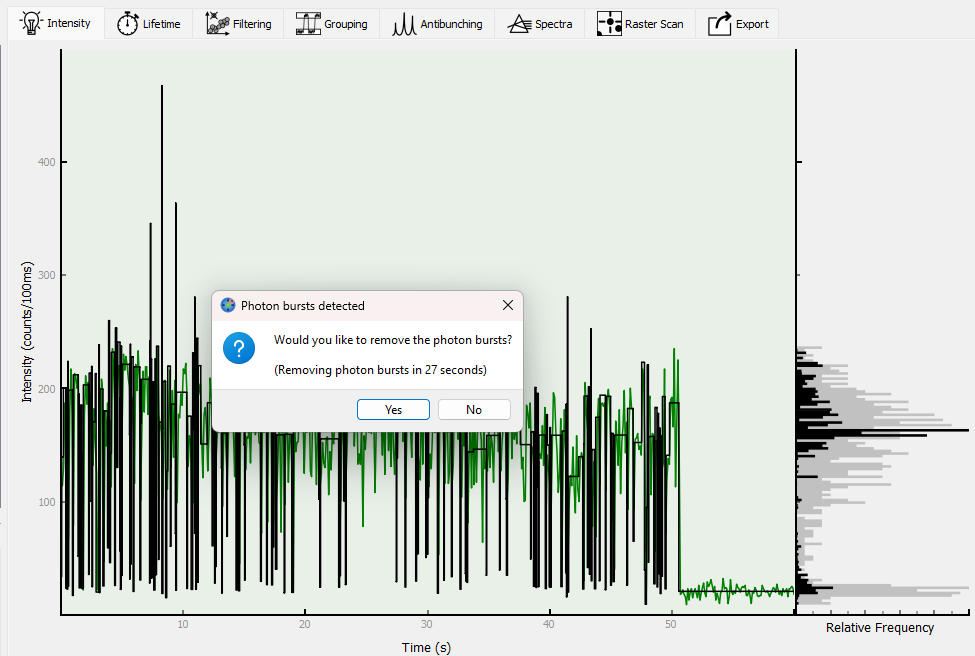}
    \vspace*{2mm}
    \caption{Photon burst dialog example with the Alexa intensity trace shown in 
    Fig.~1. This dialog is presented automatically after resolving 
    levels, if photon bursts are detected.}
    \label{suppfig:gui_photon_burst}
\end{suppfigure}

\end{document}